\documentclass[nofootinbib,aps,letterpaper,superscriptaddress,showpacs]{revtex4}
\usepackage{amsmath}
\usepackage{dsfont}
\usepackage[dvips]{graphicx}
\begin{document}

\newcommand\be{\begin{equation}}
\newcommand\ee{\end{equation}}
\newcommand\bea{\begin{eqnarray}}
\newcommand\eea{\end{eqnarray}}
\newcommand\bseq{\begin{subequations}} 
\newcommand\eseq{\end{subequations}}
\newcommand\bcas{\begin{cases}}
\newcommand\ecas{\end{cases}}
\newcommand{\p}{\partial}
\newcommand{\f}{\frac}

\title{Polymer Quantum Dynamics of the Taub Universe}

\author{Marco Valerio Battisti}
\email{battisti@icra.it}
\affiliation{Dipartimento di Fisica (G9) and ICRA,  ``Sapienza'' Universit\`a di Roma P.le A. Moro 5, 00185 Rome, Italy}
\author{Orchidea Maria Lecian}
\email{lecian@icra.it}
\affiliation{Dipartimento di Fisica (G9) and ICRA,  ``Sapienza'' Universit\`a di Roma P.le A. Moro 5, 00185 Rome, Italy}
\author{Giovanni Montani}
\email{montani@icra.it} 
\affiliation{Dipartimento di Fisica (G9) and ICRA,  ``Sapienza'' Universit\`a di Roma P.le A. Moro 5, 00185 Rome, Italy}
\affiliation{ENEA C.R. Frascati (Dipartimento F.P.N.), Via Enrico Fermi 45, 00044 Frascati, Rome, Italy}
\affiliation{ICRANET C.C. Pescara, P.le della Repubblica 10, 65100 Pescara, Italy}


\begin{abstract}
Within the framework of non-standard (Weyl) representations of the canonical commutation relations, we investigate the polymer quantization of the Taub cosmological model. The Taub model is analyzed within the Arnowitt-Deser-Misner reduction of its dynamics, by which a time variable arises. While the energy variable and its conjugate momentum are treated as ordinary Heisenberg operators, the anisotropy variable and its conjugate momentum are represented by the polymer technique. The model is analyzed at both classical and quantum level. As a result, classical trajectories flatten with respect to the potential wall, and the cosmological singularity is not probabilistically removed. In fact, the dynamics of the wave packets is characterized by an interference phenomenon, which, however, is not able to stop the evolution towards the classical singularity.  
  
\end{abstract}


\maketitle 
\section{Introduction}
The necessity for a quantum theory of gravity arises from fundamental considerations, and, in particular, from the space-time singularity problem. In fact, the classical theory of gravity implies the well known singularity theorems, among which the cosmological one \cite{haw}. The canonical quantization of gravity, which exhibits a host of difficulties both at technical and interpretative levels, is based on the Heisenberg representation of the Weyl algebra \cite{ish}. On the other hand, the background-independent formulation of canonical quantum gravity based on Yang-Mills formalism has recently appeared \cite{asht}. Anyhow, smearing such variables in the holonmy-flux representation is an important step towards canonical quantum gravity \cite{thie}. The scenario induced by such an algebra is illustrated to be equivalent to the so-called polymer representation of quantum mechanics \cite{poly,cori1,cori2}, as soon as a mechanical system is taken into account.\\
This work is aimed at investigating the quantization of the Taub model in the polymer representation of quantum mechanics.\\
The Taub Universe arises as a particular case of the Bianchi IX model, i.e. the most general scheme allowed by the homogeneity constraint  \cite{RS}. In the Bianchi IX model, the Universe dynamics towards the classical singularity is summarized by the chaotic motion of a particle. More precisely, this particle bounces an infinite number of times against the potential walls of a triangular domain, on a two-dimensional plane. The two-dimensional plane describes the configuration space of the particle (Universe) dynamics. The Taub model consists in restricting the dynamics to that of a one-dimensional particle bouncing against a wall,when only one degree of freedom is taken into account.\\
The relevance of the Taub universe in quantum cosmology is due to the fact that it is a necessary step towards the more general Bianchi IX model. The advantage of this model is that it is a generalization of other isotropic models. In particular, it has been used to test the validity of the minisuperspace scheme \cite{kuch} and to explore the application of the extrinsic cosmological time \cite{altri}. Furthermore, the Taub model has also been investigated within the framework of a generalized uncertainty principle in \cite{batt}, where the cosmological singularity has been shown to be probabilistically removed.\\ 
The polymer representation of quantum mechanics is based on a non-standard representation of the canonical commutation relations \cite{cori1}. In particular, in a two-dimensional phase space, it is possible to choose a discretized operator, whose conjugate variable cannot be promoted as an operator directly. From a physical point of view, this scheme can be interpreted as the quantum-mechanical framework for the introduction of a cutoff. Its continuum limit, which corresponds to the removal of the cutoff, has to be understood as the equivalence of  microscopically-modified theories at different scales \cite{cori2}.\\
This approach is relevant in treating the quantum-mechanical properties of a background-independent canonical quantization of gravity. In fact, the holonomy-flux algebra used in Loop Quantum Gravity reduces to a polymer-like algebra, when a system with a finite number of degrees of freedom is taken into account \cite{poly}. From a quantum-field theoretical point of view, this is substantially equivalent to introducing a lattice structure on the space \cite{polqft}. Loop Quantum Cosmology \cite{oggi} can be regarded as the implementation of this quantization technique in the minisuperspace dynamics \cite{citcor}.\\
The Taub model is approached in the scheme of an Arnowitt-Deser-Misner (ADM) reduction of the dynamics in the Poincar\'e plane. As a result, a time variable naturally emerges, and the Universe is described by an anisotropy-like variable. The anisotropy variable and its conjugate momentum are quantized within the framework of the polymer representation. More precisely, the former appears as discretized, while the latter cannot be implemented as an operator in an appropriate Hilbert space directly, but only its exponentiated version exists. The analysis is performed at both classical and quantum levels. The modifications induced by the cutoff scale on ordinary trajectories are analyzed from a classical point of view. On the other hand, the quantum regime is explored in detail by the investigation of the evolution of the wave packets of the universe.\\
Two main conclusions can be inferred. 
\begin{itemize}
	\item An interference between the wave packets and the potential wall appears. Nevertheless, the classical cosmological singularity is not probabilistically removed. In fact, the wave function of the universe is not strictly localized away from it, and the wave packets fall into it following a classical trajectory. 
	\item The comparison between the polymer approach and the Generalized Uncertainty Principle (GUP) model illustrates that the corresponding interference phenomena are produced in a complementary way. This feature appears both at classical level, as it is immediately recognized analyzing the modifications of the equations of motion, and in the quantum regime, as the behavior of the wave packets is investigated.  
\end{itemize}
The paper is organized as follows.\\
In Section II, we review the main features of the Taub cosmological model.\\
In Section III, the polymer representation of quantum mechanics is developed at both kinematic and dynamical level. Furthermore, the continuum limit of this approach is discussed.\\
The fourth Section is devoted to the application of the polymer paradigm to the Taub universe, at both classical and quantum level.\\
Section V is aimed at constructing suitable wave packets and investigating their dynamics. In particular, they are analyzed in the WDW and the polymer representation.\\
In Section VI, our results are discussed and compared with other models.\\
Concluding remarks follow.\\
Throughout the paper, we have adopted natural units, i.e. $\hbar=c=16\pi G=1$.

\section{The Taub model} 

Homogeneity reduces the configuration space of General Relativity to three dimensions. The homogeneous cosmological models \cite{RS}, the Bianchi Universes, are such that the symmetry group acts {\it simply transitively}\footnote{Let $G$ a Lie group, $G$ is said to act {\it simply transitively} on the spatial manifold $\Sigma$ if, for all $p,q\in\Sigma$, there is a unique element $g\in G$ such that $g(p)=q$.} on each spatial manifold. The Bianchi IX model, together with Bianchi VIII, is the most general one and its line element reads, in the Misner parametrization \cite{Mis69}, 
\be 
ds^2=N^2dt^2-e^{2\alpha}\left(e^{2\gamma}\right)_{ij}\omega^i\otimes\omega^j, 
\ee 
where $N=N(t)$ is the lapse function and the left invariant 1-forms $\omega^i=\omega^i_adx^a$ satisfy the Maurer-Cartan equation $2d\omega^i=\epsilon^i_{jk}\omega^j\wedge\omega^k$. The variable $\alpha=\alpha(t)$ describes the isotropic expansion of the Universe and $\gamma_{ij}=\gamma_{ij}(t)$ is a traceless symmetric matrix, $\gamma_{ij}=diag\left(\gamma_++\sqrt3\gamma_-,\gamma_+-\sqrt3\gamma_-,-2\gamma_+\right)$, which determines the anisotropy changes via $\gamma_\pm$. The classical singularity appears for $\alpha\rightarrow-\infty$, since the determinant of the 3-metric is given by $h=\det e^{\alpha+\gamma_{ij}}=e^{3\alpha}$.

\paragraph{Canonical Analysis.} The Hamiltonian constraint for this model is obtained performing the usual Legendre transformation. As well-known \cite{Mis69,CGM}, the dynamics of the Universe towards the singularity is described by the motion of a two-dimensional particle (the two physical degrees of freedom of the gravitational field) in a dynamically-closed domain. Such a domain depends on the time variable $\alpha$ in the Misner picture, while it is stationary in the Misner-Chitr\'e framework defined by the variables \cite{Chi}
\be\label{txt} 
\alpha=-e^\tau\xi, \qquad \gamma_+=e^\tau\sqrt{\xi^2-1}\cos\theta, \qquad \gamma_-=e^\tau\sqrt{\xi^2-1}\sin\theta, 
\ee 
with $\xi\in[1,\infty)$ and $\theta\in[0,2\pi]$. In fact, the dynamically-allowed domain becomes independent of $\tau$, which behaves like a time variable. In terms of these new variables, the Hamiltonian constraint rewrites  
\be 
H=-p_\tau^2+p_\xi^2(\xi^2-1)+\f{p_\theta^2}{\xi^2-1}\approx0. 
\ee 
\begin{figure} 
\begin{center} 
\includegraphics[height=2in]{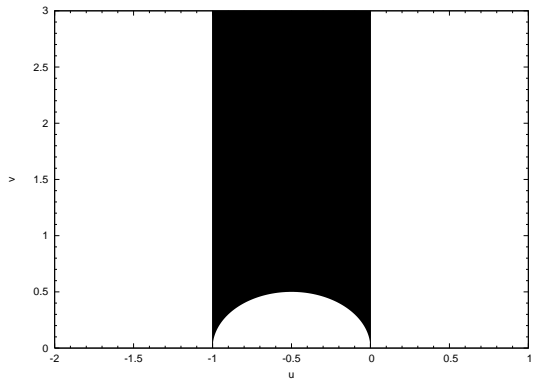} 
\caption{The dynamical-allowed domain $\Gamma_Q(u,v)$ in the Poincar\'e complex upper half-plane where the dynamics of the Universe is restricted, towards the classical singularity, by the potential.} 
\end{center} 
\end{figure} 

\paragraph{ADM Reduction.} Let us perform the ADM \cite{adm} reduction of the dynamics. This scheme relies on the idea to solve the classical constraint with respect to a given momentum, before implementing any quantization algorithm. This paradigm allows us to dynamically separate the six-dimensional phase space of the model. In particular, a time variable arises and an effective Hamiltonian, which will depend only on the physical degrees of freedom of the system (the anisotropy-like variables), comes out. We solve explicitly the constraint $H=0$ with respect to $p_\tau$, and thus we consider the variable $\tau$ as the time coordinate for the dynamics (we adopt the time gauge $\dot\tau=1$), obtaining
\be\label{hxt} 
-p_\tau=\sqrt{p_\xi^2(\xi^2-1)+\f{p_\theta^2}{\xi^2-1}}. 
\ee 
The dynamics of such a system is equivalent to a billiard ball on a Lobatchevsky plane \cite{ChBa83}, as we can see by means of the Jacobi metric\footnote{This approach reduces the equations of motion of a generic system to a geodesic problem on a given manifold.}. It is possible to choose the so-called Poincar\'e representation in the complex upper half-plane \cite{KM97} by using new variables $(u,v)$, defined as 
\be\label{uv} 
\xi=\f{1+u+u^2+v^2}{\sqrt3v}, \qquad \theta=-\tan^{-1}\left(\f{\sqrt3(1+2u)}{-1+2u+2u^2+2v^2}\right). 
\ee 
The dynamical-allowed domain $\Gamma_Q=\Gamma_Q(u,v)$ is plotted in Fig. 1. It is worth noting that the three corners in the Misner picture are replaced by the points $(0,0)$, $(-1,0)$ and $v\rightarrow\infty$ in the $(u-v)$ plane. In this scheme, the ADM ``constraint'' is simpler than the previous one (\ref{hxt}), and becomes
\be\label{huv} 
-p_\tau\equiv H_{ADM}=v\sqrt{p_u^2+p_v^2}. 
\ee 
The Taub Universe corresponds to the Bianchi IX one in the particular case of $\gamma_-=0$ \cite{RS}. The phase space of this model is four-dimensional and its dynamics is equivalent to the motion of a particle in a one-dimensional domain. Considering such a domain corresponds to taking only one of the three equivalent potential walls of the Bianchi IX model. As we can see from (\ref{txt}) and (\ref{uv}), this particular case appears for $\theta=0\Rightarrow u=-1/2$ ($\xi=(v^2+3/4)/\sqrt3v$), and the ADM Hamiltonian (\ref{huv}) rewrites 
\be\label{hv} 
H_{ADM}^T=vp_v, 
\ee 
being $v\in[1/2,\infty)$, as shown in Fig. 1. The Hamiltonian above (\ref{hv}) can be further simplified defining a new variable $x=\ln v$, and becomes
\be\label{ht} 
H_{ADM}^T=p_x\equiv p, 
\ee 
which will be the starting point of our analysis. Within this framework, the Taub model is therefore described by a two-dimensional system in which the variable $\tau$ is considered as the time, while the variable $x$ describes the single degree of freedom of the Universe, i.e. the shape change. It is worth stressing that the classical singularity now appears for $\tau\rightarrow\infty$.

\section{Polymer Quantum Mechanics}

The polymer representation of quantum mechanics consists in defining abstract kets, labeled by a real number, and then considering a suitable finite subset of them, whose Hilbert space is defined by the corresponding inner product \cite{cori1}. This procedure can be shown to be an inequivalent representation of the Weyl algebra wrt the ordinary Schroedinger one. This representation helps one gain insight onto some particular features of quantum mechanics, when an underlying discrete structure is somehow hypothesized. The request that the Hamiltonian associated to the system be of direct physical interpretation defines the polymer phase space, and the continuum limit can be recovered by the introduction of the concept of scale \cite{cori2}.\\
 
\paragraph{Wave functions and operators.}
One can start by considering abstract kets $|\mu>$, $\mu\in\mathds{R}$, and a suitable subset defined by $\mu_i\in\mathds{R}$, $i=1,2,..N$. These kets are assumed to be an orthonormal basis, i.e., $<\mu|\nu>=\delta_{\mu\nu}$, along which any state $\phi$ can be projected. This defines a Hilbert space $\mathcal{H}_{pol}$, on which two basic operators act, the symmetric ''label'' operator, $\hat{\epsilon}$, such that $\hat{\epsilon}|\mu>=\mu|\mu>$, and a one-parameter family of unitary operators, $\hat{s}(\lambda)$, such that $\hat{s}(\lambda)|\mu>=|\mu+\lambda>$. Because all kets are orthonormal, $\hat{s}(\lambda)$ is discontinuous, and cannot be obtained from any Hermitian operator by exponentiation. It is worth noting that this Hilbert space is not separable\footnote{A Hilbert space is separable if and only if it admits a countable orthonormal basis.}.\\
For the toy model of a 1-dimensional system, whose phase space is described by the variables $p$ and $q$, the polymer representation techniques find interesting applications when one of the two variables is supposed to be discrete. This discreteness will affect both wave functions, obtained by projecting the physical state on the $p$ or $q$ basis (polarization), and the operators associated to the canonical variables, acting on them.\\
For later purposes, we will discuss only the case of a discrete position variable $q$, and the corresponding momentum polarization.\\
In this case, wave functions are given by $\psi_{\mu}(p)=<p|\mu>=e^{ip\mu}$. Accordingly, the ''label'' operator $\hat{\epsilon}$ is easily identified with $\hat{q}$, i.e., $\hat{q}\phi_\mu=-i\partial_p\psi_\mu=\mu\psi_\mu$, while the ''shift'' operator does not exist, as discussed previously.\\
It can be shown that the corresponding Hilbert space is $\mathcal{H}_{pol}=L^2(\mathds{R}_B,d\mu_H)$, i.e. the set of square-integrable functions defined on the Bohr compactification of the real line $\mathds{R}_B$, with a Haar measure $d\mu_H$. Since the kets $|\mu>$ are arbitrary but finite, the wave functions can be interpreted as a quasi-periodic function, with the inner product
\begin{equation}\label{inpro}
<\psi_\mu|\psi_\lambda>=\int_{\mathds{R}_B}d\mu_H\bar{\psi}_\mu(p)\psi_\lambda(p)=\lim_{L\rightarrow\infty}\frac{1}{2L}\int_{-L}^{L}dp\bar{\psi}_\mu(p)\psi_\lambda(p)=\delta_{\mu,\lambda}.
\end{equation}
 
\paragraph{Dynamics.}
The Hamiltonian operator $H$ describing a quantum-mechanical system is usually a function of both coordinate an momentum, i.e. $H=H(q,p)=\frac{p^2}{2m}+V(q)$ while, in the particular case of a discrete position variable in the momentum polarization, $p$ cannot be implemented as an operator, so that some restrictions on the model have to be required.\\
As a first step, a suitable approximation for the kinetic term has to be provided. For this purpose, it is useful to restrict the arbitrary kets $|\mu_i>$, $i\in\mathds{R}$ to $|\mu_i>$, $i\in\mathds{Z}$, i.e. to introduce the notion of regular graph $\gamma_{\mu_0}$, defined as a numerable set of equidistant points, whose separation is given by the parameter $\mu_0$, $\gamma_{\mu_0}=\left\{q\in\mathds{R}|q=n\mu_0, \forall n\in\mathds{Z}\right\}$. The associated Hilbert space $\mathcal{H}_{\gamma_{\mu_0}}$ is separable. Because of the regular graph $\mu_0$, the eigenfunctions of $\hat{p}_{\mu_0}$ must be of the form $e^{im\mu_0 p}$, $m\in\mathds{Z}$, which are Fourier modes, of period $2\pi/\mu_0$. The inner product (\ref{inpro}) is equivalent to the inner product on a circle $S^1$ with uniform measure, i.e.,
\begin{equation}
<\phi(p)|\psi(p)>_{\mu_0}=\frac{\mu_0}{2\pi}\int^{\pi/\mu_0}_{-\pi/\mu_0}\hat{\phi}(p)\psi(p),
\end{equation}
with $p\in\left(-\pi\mu_0, \pi/\mu_0\right)$, so that $\mathcal{H}_{\gamma_{\mu_0}}=L^2(S^1,dp)$.
Within this space, it is possible to construct an approximation for the ''shift'' operator, i.e. a regulated operator $\hat{p}_{\mu_0}$,
\begin{equation}\label{incre}
\hat{p}_{\mu_0}|\mu_n>=\frac{i}{2\mu_0}\left(|\mu_{n+1}>-|\mu_{n-1}>\right).
\end{equation}
More precisely, the polymer paradigm can be understood as the formal substitution 
\begin{equation}\label{para}
p\rightarrow \frac{1}{\mu_0}\sin (\mu_0p),
\end{equation}
where the incremental ratio (\ref{incre}) has been evaluated for exponentiated operators. The Hamiltonian operator $H_{\mu_0}$, which lives in $\mathcal{H}_{\gamma_{\mu_0}}$, reads $H_{\mu_0}=\frac{\hat{p}_{\mu_0}^2}{2m}+V(\hat{q})$, where the action of the new multiplication operator $\hat{p}_{\mu_0}$ on wave functions in the momentum polarization is 
\begin{equation}\label{mom}
\hat{p}_{\mu_0}^2\psi(p)=\frac{2}{\mu_0^2}\left[1-\cos(p\mu_0)\right],
\end{equation}
while the differential operator $q$ is well defined.\\

\paragraph{Continuum Limit.}\label{conti}
The physical Hilbert space of such theories can be constructed as the continuum limit of effective theories at different scales, and can be illustrated to be unitarily isomorphic to the ordinary one, $\mathcal{H}_S=L^2(\mathds{R}, dp)$.\\
To this end, it is useful to remark that it is impossible to obtain $\mathcal{H}_S$ starting from a given graph $\gamma_0=\left\{ q_k\in\mathds{R}|q_k=ka_0, \forall k\in\mathds{Z}\right\}$ by dividing each interval $a_0$ into $2^n$ in new intervals of length $a_n=a_0/2^n$, because $\mathcal{H}_S$ cannot be embedded into $\mathcal{H}_{pol}$.\\
It is however possible to go the other way round and to look for a continuous wave function that is approximated by a wave function over a graph, in the limit of the graph becoming finer. In fact, if one defines a scale $C_n$, i.e., a decomposition of $\mathds{R}$ in terms of the union of closed-open intervals that have lattice points as end points and cover $\mathds{R}$ without intersecting, one is then able to approximate continuous functions with functions that are constant on these intervals. As a result, at any given scale $C_n$, the kinetic term of the Hamiltonian operator can be approximated as in (\ref{mom}), and effective theories at given scales are related by coarse-graining maps. In particular, it is necessary to regularize the Hamiltonian, treated as a quadratic form, as a self-adjoint operator at each scale by introducing a normalization factor in the inner product. The convergence of microscopically-corrected Hamiltonians is based on the convergence of energy levels and on the existence of completely normalized eigencoverctors compatible with the coarse-graining operation.  

\section{Polymer Taub Universe}

In this section, we will apply the polymer discretization technique to the description of the Taub model. In particular, we will specify the Hamiltonian (\ref{ht}) for the case of a discretized $x$ space. As a result, the conjugate variable will not be implemented to operator directly, in the corresponding Hilbert space. Furthermore, the momentum space will be compactified, the compactification scale depending on the lattice characteristic length.\\
The modifications to the Taub universe induced by the polymer representation will be investigated at both classical and quantum level.

\subsection{Classical Analysis}
First of all, let us clarify the physical meaning of our variables. The configuration variable $x$ is related to the Universe anisotropy $\gamma_+$ via the expression (\ref{txt}), for $\theta=0$ and $\xi=(v^2+3/4)/\sqrt3v$, as 
\be\label{anix}
\gamma_+=\f{e^\tau}{\sqrt3v}\left(v^2-\f34\right)=\f{e^{\tau-x}}{\sqrt3}\left(e^{2x}-\f34\right).
\ee
By this equation, a monotonic relation between the anisotropy of the Universe $\gamma_+$ and our (classical) configuration variable $x=\ln v\in[x_0\equiv\ln(1/2),\infty)$ appears, and, therefore, the variable $x$ can be regarded as a measure of the model anisotropy. In particular, the isotropic shape of the Taub Universe ($\gamma_+=0$) comes out for a particular value of $x$, i.e. $x=\ln(\sqrt3/2)$, and, in this case, we get the closed Friedmann-Robertson-Walker Universe.\\
Let us now discuss the polymer dynamics of a Taub universe at classical level. By means of the substitution (\ref{para}), the Taub Hamiltonian reads
\be\label{gatto}
H=\frac{1}{a_n}\sin (a_n p).
\ee 
From now on, we will take into account the discussion about the definition of a scale, and, for the sake of compact notation, we will drop the index $n$ from $a_n$. From a classical point of view, the equations of motion are
\begin{subequations}
\begin{align}
&\dot{x}=\left\{x,H\right\}=\cos(ap),\\
&\dot{p}=\left\{p,H\right\}=0,
\end{align}
\end{subequations}
where dot denotes differentiation with respect to the time variable $\tau$. The equations of motion are immediately solved as
\begin{subequations}\label{eqmot}
\begin{align}
&x(\tau)=\cos(ap) \tau,\\
&p(\tau)=A,
\end{align}
\end{subequations}
where $A$ is a constant.\\
As well understood, the system (\ref{eqmot}) describes a free particle (Universe) bouncing against a wall.\\
In the ordinary case, i.e. for $a=0$, the model can be interpreted as a photon in the Lorentzian minisuperspace, and the classical trajectory in the $(\tau-x)$ plane is its light-cone. More precisely, the incoming particle ($\tau<0$) bounces on the wall ($x=x_0$) and falls into the classical cosmological singularity ($\tau\rightarrow\infty$).\\
Contrastingly, in the discretized case, i.e. for $a\neq0$, the one-parameter family of trajectories flattens, i.e. the angle between the incoming trajectory and the outgoing one is greater than $\pi/2$ since $p\in\left(-\pi/a, \pi/a\right)$ (see Fig. 2). As these trajectories diverge rather than converging, we expect the polymer quantum effects to be reduced with respect to the classical case, as we will verify below.
\begin{figure}
\begin{center}
\includegraphics[height=2in]{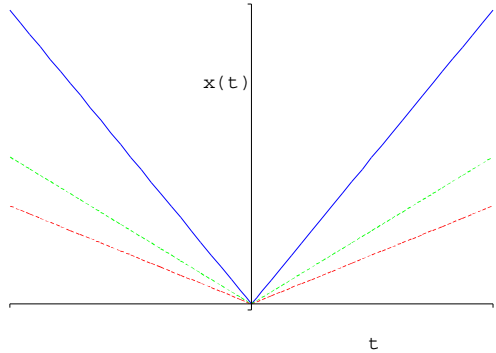}
\caption{Semiclassical equations of motion for the Taub Universe: ordinary trajectory (blue line, $a=0$) and polymer trajectories (green ($\cos aA=1/2$) and red ($\cos aA=1/3$) dashed lines).} 
\end{center}
\end{figure}

\subsection{Quantum Regime}
We now investigate the quantum behavior of the model. After analyzing the mathematical requirements of the polymer representation and their physical implications for the model, we apply the methods introduced above to the Taub Universe. In particular, we choose a discretized $x$
space, and solve the corresponding eigenvalue problem in the $p$ polarization. \\
Even though the bulk of the discussion of the relation of the polymer representation at different scales is based on the properties of the Hamiltonian as a quadratic form, we can nevertheless apply this paradigm to the Taub model, which is described by a linear Hamiltonian (\ref{ht}), after the well-known procedure, established in \cite{Puzio}. In fact, squaring the Hamiltonian leads to squared eigenvalues without affecting the corresponding eigenfunctions.\\
 
We are now ready to analyze the Schroedinger equation $i\partial_\tau\Psi=p\Psi$ for the wave function $\Psi=\Psi(p,\tau)$ corresponding to (\ref{ht}), where the configuation variable $x$ is defined in the domain $x\in[x_0\equiv\ln(1/2),\infty)$.\\
Considering the time evolution for the wave function $\Psi$ as given by $\Psi_k(p,\tau)=e^{-ik\tau}\psi_k(p)$ and the results of \cite{Puzio}, we obtain the following eigenvalue problem
\be
(p^2-k^2)\psi_k(p)=\left[\frac{2}{a^2}\left(1-\cos(ap)\right)-k^2\right]\psi_k(p),
\ee 
where, in the last step, the substitution (\ref{mom}) has been taken into account. This eigenvalue problem is solved by
\begin{subequations}
\begin{align}
&k^2=k^2(a)=\frac{2}{a^2}\left(1-\cos(ap)\right)\leq k^2_{max}=\frac{4}{a^2}\label{kappa}\\
&\psi_{k,a}(p)=A\delta(p-p_{k,a})+B\delta(p+p_{k,a})\label{pi}\\
&\psi_{k,a}(x)=A\left[\exp(ip_{k,a} x)-\exp(ip_{k,a}(2x_0-x))\right]\label{ics}:
\end{align}
\end{subequations}
(\ref{pi}) is the momentum wave function, with $A$ and $B$ two arbitrary integration constant, and (\ref{ics}) is the coordinate wave function, where an integration constant has been eliminated by imposing suitable boundary conditions. Moreover, we have defined the modified dispersion relation
\be\label{disprel}
p_{k,a}\equiv\frac{1}{a}\arccos\left(1-\frac{k^2a^2}{2}\right)
\ee 
from (\ref{kappa}). Furthermore, we stress that $k^2$ is bounded from above, as illustrated in (\ref{kappa}), but it is its square root, considered for its positive determination, which accounts for the time evolution of the wave function.

\section{Taub Wavepackets}

We will now gain insight onto the physical implications of the model by constructing suitable wave packets $\Psi(x,\tau)$. In fact, analyzing the dynamics of such wave packets allows us to give a precise description of the evolution of the Taub model. Such an evolution will be preformed in both the polymer and  Wheeler-DeWitt (WDW) approaches. More precisely, the latter will be considered the proper continuum limit of the polymer representation, as illustrated above. Wavepackets are a superposition of eigenfunctions (\ref{ics}), such as
\be\label{wp}
\Psi(x,\tau)=\int_0^{k_{max}} dk A(k)\psi_{k,a}(x)e^{-ik\tau},
\ee 
where $A(k)$ is a Gaussian weighting function, i.e. $A(k)=\exp[-(k-k_0)^2/2\sigma^2]$.\\

\subsection{WDW Dynamics}

To better understand the modifications induced on the ordinary dynamics by the polymer representation, we briefly summarize the WDW wave packet dynamics for the Taub model. In this case, the Hamiltonian is simply (\ref{ht}); the associated Schroedinger eigenvalue equation can be solved directly: the eigenfunctions in the position representation are just plane waves. This way, wave packets (\ref{wp}) can be analytically calculated, with no upper limit for the energy $k$. The result is plotted in Fig. 3. As we can see from the picture, the wave packets follow the ordinary classical trajectories described in the previous Section. The probability amplitude to find the particle (Universe) is peaked around these trajectories. In this respect, no privileged regions arise, namely no dominant probability peaks appear in the ($\tau-x$) plane. As a matter of fact, the ``incoming'' Universe ($\tau<0$) bounces at the potential wall ($x=x_0$) and then falls towards the classical singularity ($\tau\rightarrow\infty$). Therefore, as well-known, the WDW formalism is not able to shed light on the necessary quantum resolution of the classical cosmological singularity. As we will see below, this picture is slightly modified in the polymer representation.   
\begin{figure}
\begin{center}
\includegraphics[height=2in]{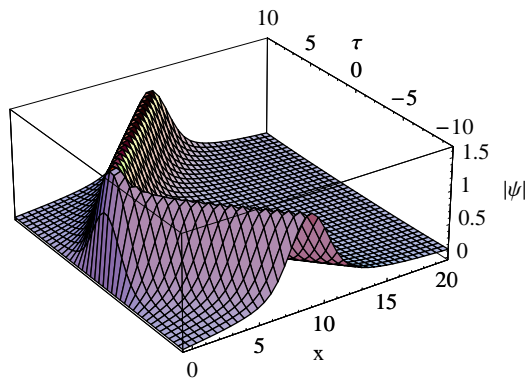}
\caption{The WDW wave packet $\mid\Psi(x,\tau)\mid$ for the Taub model, i.e. $a=0$ ($k_0=0.1$, $\sigma=1$).} 
\end{center}
\end{figure}

\subsection{Polymer Dynamics}

We are now ready to analyze the modifications brought by the polymer representation in the quantized Taub Universe. Two cases can be distinguished, i.e the case $k_0a\sim\mathcal{O}(1)$, for which it is not possible to recover the ordinary representation of the momentum operator, and the case $k_0a\ll1$, for which such a treatment is feasible. For $k_0a\sim\mathcal{O}(1)$, we get remarkable modifications of the wave packet evolution. From a probabilistic point of view, however, such modifications do not remove the cosmological singularity. The case $k_0a\ll1$, contrastingly, can be considered as the semiclassical limit of the polymer approach.\\

\paragraph{Peaked Weighting function.}
Let us now investigate the first case, $k_0a\sim1$, where the implementation of the polymer substitution (\ref{para}) does not lead to the ordinary Schroedinger dynamics. Furthermore, we stress that the choice of the  value for the standard deviation $\sigma$ in the Gaussian weighting function can be relevant for detecting the effects of the polymer paradigm.\\
In fact, if the weighting function is very sharply peaked around any value $k_0$, the resulting wave packet will be well-approximated by a purely monochromatic wave, for which a narrow neighborhood of $k_0$ is selected. As a consequence, the ordinary dispersion relation is effectively reproduced by the deformed one, (\ref{disprel}). In fact, narrowing the range of $k$ is equivalent to expand the deformed Hamiltonian (\ref{gatto}) around a given value of the momentum. This kind of behavior is explicitly illustrated in Fig. 4, where it is possible to appreciate a small interference phenomenon between the incoming (outgoing) wave and the wall. This feature can be interpreted as a relic of the polymer modifications of the Taub Universe dynamics, as it will be clearer in the next analysis.\\
\begin{figure}
\begin{center}
\includegraphics[height=2in]{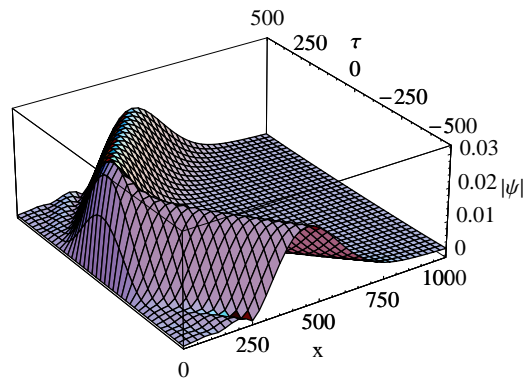}
\caption{The peaked polymer wave packet $\mid\Psi(x,\tau)\mid$ for the Taub model, with $k_0 a=1/2$ ($a=50$, $k_0=0.01$, $\sigma=0.0125$).} 
\end{center}
\end{figure} 

\paragraph{Spread Weighting function.}
On the basis of the previous analysis, the effects of the polymer substitution show up when broad wave packets are considered, i.e. when a large neighborhood of $k_0$ is taken into account by the Gaussian weighting function.\\
In this case, it is possible to appreciate all the modifications induced by the deformed Hamiltonian (\ref{gatto}). As a result, a strong interference phenomenon appears between the incoming (outgoing) wave and the wall. However, as a matter of fact, such an interference phenomenon is not able to localize the wave packet in a determined region of the configuration space. This way, the probability density to find the Universe far away the singularity is not peaked, i.e. the cosmological singularity of this model is not tamed by the polymer representation from a probabilistic point of view. Consequently, the incoming particle (Universe) is initially ($t<0$) localized around the classical polymer trajectory (\ref{eqmot}). It then bounces against the wall ($x=x_0$), where the wave packet spreads in the ''outer'' region, regains the classical polymer trajectory ($t>0$) and eventually falls into the cosmological singularity ($t\rightarrow\infty$). This way, we claim that the classical singularity is not solved by this quantization of the model.\\
It is interesting to remark that the interference phenomenon occurs in the ''outer'' region of the configuration space, the ($\tau-x$) plane. These features are explained in Fig. 5. As we will discuss later on, such a behavior is complementary to that observed in the case of a generalized uncertainty principle.
\begin{figure}
\begin{center}
\includegraphics[height=2in]{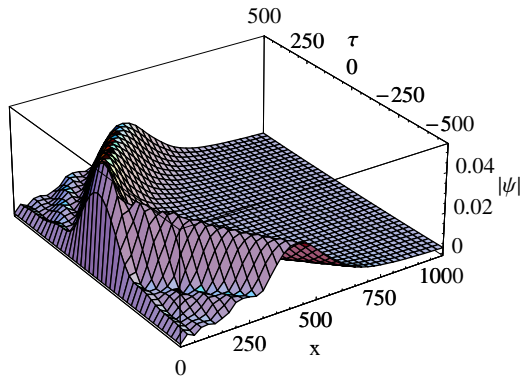}
\caption{The spread polymer wave packet $\mid\Psi(x,\tau)\mid$ for the Taub model, with $k_0 a=1/2$ ($a=50$, $k_0=0.01$, $\sigma=0.125$).} 
\end{center}
\end{figure}

\subsection{Semiclassical Limit}
We end up our analysis by obtaining the correct semiclassical limit of the model. Within this framework, to obtain the proper continuum limit of the polymer representation, the value of $k_0$ is not arbitrary, but has to be chosen according to the request $k_0a\ll1$. Since the range of the variable conjugated to the anisotropy variable is compactified, then  $k_0$ has to be small with respect to the length of the interval\footnote{ We recall that the length of the integration interval $L$ of (\ref{wp}) is $L\propto1/a$, so that $k_0\ll L$.}.\\
As a result, differently from the other cases, the value of $k_0$ around which the wave packet is peaked is not arbitrary, but constrained by the characteristic scale $a$ we are investigating. The ordinary WDW behavior is therefore recast, as plotted in Fig. 6. Even though taking $ap\ll1$ is enough to reproduce the ordinary Hamiltonian (as a general feature of the polymer representation because of relation (\ref{para})), the fact that the correct semiclassical limit for the polymer quantum Taub Universe is obtained for a wave packet peaked at $k_0\ll1/a$ is a non-trivial feature of the model.
\begin{figure}
\begin{center}
\includegraphics[height=2in]{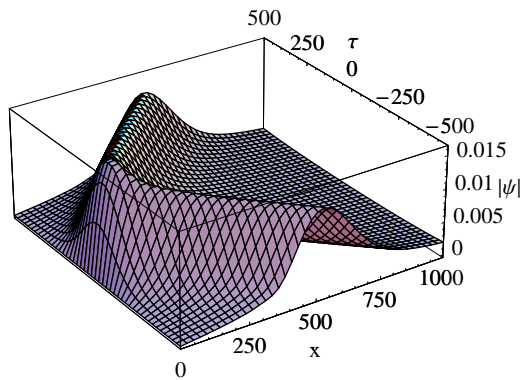}
\caption{The semiclassical limit of the polymer wave packet $\mid\Psi(x,\tau)\mid$ for the Taub model, with $k_0 a=1/20$ ($a=50$, $k_0=0.001$, $\sigma=0.01$).} 
\end{center}
\end{figure}

\section{Comparison with other approaches}
We can deeper understand the physical implications of this model by comparing it with other applications of the polymer representation in cosmology and with the implementation of a generalized uncertainty principle for the Taub Universe. In fact, in our model, the cosmological singularity is not probabilistically suppressed, as one could expect from other models. Let us now discuss the main differences from those models.\\

\paragraph{Isotropic Polymer Cosmology.} The fact that the cosmological singularity is not removed within this framework could look apparently in contrast with other models, such as \cite{bb,citcor}: in the cosmological isotropic sector of General Relativity, i.e. the FRW models, the singularity is removed by loop quantum effects. In particular, the wave function of the universe exhibits a non-singular behavior at the classical singularity, and the Big Bang is replaced by a Big Bounce, when a free scalar field is taken as the relational time \cite{qn}. There are however at least two fundamental differences with respect to our model.
\begin{itemize}
	\item Within our scheme, the variable $\tau$, which describes the isotropic expansion of the Universe, is not discretized, but treated in the ordinary way. In fact, in the ADM reduction of the model, this variable emerges as the time coordinate, and cannot be discretized in a polymer frmework. More precisely, the phase space of this model is four-dimensional, but we naturally select a two-dimensional submanifold of it, i.e. the ($x-p_x$) plane, where we implement the polymer paradigm. In other words, we must discretize the anisotropy variable only, without modifying the volume (time) one. On the other hand, in the FRW case, the scale factor of the universe is directly quantized by the use of the polymer (loop) techniques. So far, the evolution itself of the wave packet of the universe is deeply modified by such an approach.
	\item The solution of the equations of motion is radically different in the two cases. In fact, in our case, the variable $p$, conjugated to the anisotropy, is a constant of motion, and, from the Schroedinger equation, it describes also $k$, the energy of the system. According to the polymer substitution (\ref{para}), it is always possible to choose a scale $a$ for which the polymer effects are negligible during the whole dynamics, at  classical level. On the other hand, the Hamiltonian constraint in the FRW case does not allow for a constant solution of the variable conjugated to the scale factor. For this reason, it is not possible to choose a scale, such that the polymer modifications are negligible throughout the whole evolution.\\
   
\end{itemize}

\paragraph{Homogeneous Loop Cosmology.} Also the Bianchi cosmological models have been analyzed in the framework of Loop Quantum Cosmology, according to the ADM reduction of the dynamics. The main difference between these works and our approach consists in the fact that in \cite{homo1} all the degrees of freedom are quantized by Loop techniques. In particular, also the time variable, i.e. the Universe volume, is treated at the same level as the others. In most cases, the time variable is defined by a phase space variable, i.e. it is an internal one. As a result, also the Bianchi Universes are singularity-free \cite{homo2}. In this respect, our analysis is based on considering the time variable as an ordinary Heisenberg variable.\\   

\paragraph{GUP Cosmology.} The Taub universe, in this ADM reduction, has also been described within the framework of GUP \cite{batt}. In that case, the conjugate variables $x-p_x$ are quantized by means of a deformed Heisenberg algebra. As a result, the cosmological singularity is probabilistically suppressed, since the deformation parameter helps localize the wave function of the universe far away from it. This way, comparing the GUP approach and the polymer one allows us to infer that it i not always sufficient to ''deform'' the anisotropy variable to obtain significant modifications on the universe evolution. However, the polymer paradigm is a Weyl representation of the commutation relations, while, as explained in \cite{kem}, a generalization of the commutation relations cannot by obtained by a canonical transformation of the Poisson brackets of the system.\\
Moreover, it is possible to show how the effective framework of loop cosmological dynamics can be obtained by the opposite sign of the deformation term of the modified Heisenberg algebra \cite{bdc}. This feature is phenomenologically in agreement with our analysis.

\section{Concluding remarks}  
In this work, we have analyzed the polymer quantization of the Taub Universe. The Taub model admits a four-dimensional phase space, and its ADM reduction allows for an emerging time variable. So far, the energy variable and its conjugate momenta are treated canonically, while the anisotropy variable and its conjugate momenta are quantized according to the polymer paradigm. In particular, the anisotropy variable is assumed as discrete, while its conjugate momenta is replaced by its exponentiated version on a compactified space.\\
This investigation has been developed at both classical and quantum levels. In the first case, trajectories are illustrated to flatten, with respect to the standard case. However, the most interesting result appears at the quantum level, when the evolution of wave packets is discussed. In fact, an interference phenomenon is illustrated to occur between the potential wall and the incoming particle (Universe), described as a localized wave packet. Nevertheless, the interference is not strong enough for the wave packet evolution to be localized. As a result, the corresponding outgoing particle (Universe) appears, whose evolution towards the cosmological singularity is not probabilistically avoided.\\
The features of the polymer Taub Universe enhance the comparison with other approaches. On the one hand, the polymer quantization technique has been also applied to isotropic models. In this case, the choice of the scale factor as the polymer-discrete variable involves the singularity directly. This way, a non-singular quantum cosmology arises. On the other hand, the GUP approach to the Taub model leads to a singularity-free Universe. In particular, from an effective point of view, the consequences of the polymer scheme are complementary to those predicted by the GUP framework.

\end{document}